\documentclass[twocolumn]{svjour3}          

\usepackage{graphicx}                       
\usepackage{amsmath,amsfonts,amssymb}       
\usepackage{hyperref}                       
\usepackage{float}                          
\usepackage{lipsum}                         
\usepackage{newtxtext}                      
\usepackage{newtxmath}                      
\usepackage{multirow}                       
\usepackage[title]{appendix}                
\usepackage{xcolor}                         
\usepackage{textcomp}                       
\usepackage{booktabs}                       
\usepackage{algorithm}                      
\usepackage{algpseudocode}                  
\usepackage{listings}                       
\usepackage{tcolorbox}
\journalname{Springer Journal}              
\usepackage{multirow} 
\usepackage{hhline}   
\usepackage{enumitem} 

\title{Fighting AI with AI: AI-Agent Augmented DNS Blocking of LLM Services during Student Evaluations}
\author{
        Yonas Kassa \textsuperscript{1} \and 
        James Bonacci
 \textsuperscript{1} \and
        Ping Wang \textsuperscript{1}} 
\institute{
    \textsuperscript{1}School of Data Intelligence and Technology, Robert Morris University, Pittsburgh, USA. \\
    *Corresponding author(s). E-mail(s): \href{mailto:ykassa@acm.org}{ykassa@acm.org} \\
    Contributing authors: 
    \href{mailto:ykassa@acm.org}{Yonas Kassa};
    \href{mailto:jxbst976@mail.rmu.edu}{James Bonacci};
    \href{mailto:awangp@rmu.edu}{Ping Wang}; \\
}
\date{}
\begin{document}
\maketitle
\begin{abstract}
The transformative potential of large language models (LLMs) in education, such as improving accessibility and personalized learning, is being eclipsed by significant challenges. These challenges stem from concerns that LLMs undermine academic assessment by enabling bypassing of critical thinking, leading to increased cognitive offloading. This emerging trend stresses the dual imperative of harnessing AI’s educational benefits while safeguarding critical thinking and academic rigor in the evolving AI ecosystem.
To this end, we introduce AI-Sinkhole, an AI-agent augmented DNS-based framework that dynamically discovers, semantically classifies, and temporarily network-wide blocks emerging LLM chatbot services during proctored exams. AI-Sinkhole offers explainable classification via quantized LLMs (LLama 3, DeepSeek-R1, Qwen-3) and dynamic DNS blocking with Pi-Hole. We also share our observations in using LLMs as explainable classifiers which achieved robust cross-lingual performance (F1-score $>$ 0.83). To support future research and development in this domain initial codes with a readily deployable `AI-Sinkhole' blockist is available on \textbf{https://github.com/AIMLEdu/ai-sinkhole}.
\end{abstract}


\section{Introduction}
The advent of large language models (LLMs), led by systems such as OpenAI’s ChatGPT, has fundamentally transformed digital information processing, enabling users to retrieve, summarize, and synthesize vast amounts of knowledge with unprecedented ease. These models have rapidly evolved from error-prone prototypes challenged by hallucinations and misalignment to sophisticated tools that excel in diverse benchmarks including mathematics, science, medicine and even autonomously execute tasks using external tools (e.g., browsers) with agentic AI paradigms\cite{yang2025agentic}. 
Their integration into domains such as education has also been transformative, with LLMs serving as dynamic textbooks, personalized tutors, and research assistants\cite{wen2024ai}. However, their transformative potential is increasingly being shadowed by challenges in academic evaluations\cite{cotton2024chatting}. 
Misuse and overreliance on LLMs during evaluations, such as using AI chatbots to solve exam questions, undermines the core purpose of academic assessments which is to test the knowledge and skills students have acquired while also measuring critical thinking and independent problem-solving\cite{gerlich2025ai}. Nowadays it is even difficult to stay focused without being tempted to peek into a solution generated by an AI while trying to perform organic search.
Compounding the challenge is the rapid proliferation and ubiquity of LLM chatbots in digital ecosystems, where students can now bypass learning objectives by simply pasting questions into AI chatbots, which often return complete solutions or entire source code within seconds. 
Existing AI generated content detection systems, including those developed by OpenAI or academic research, have proven inadequate in addressing these challenges\cite{weber2023testing}. Existing methods like static DNS-based blocking (e.g., domain or IP blacklists ) also fail to adapt to the rapid proliferation of new platforms and aggregator services. Aggregator services offer customized solutions such as free access to leading premium LLMs for people who cannot afford paying for multiple online services\cite{kassa2025online}. The increasing availability of open-weight models and API services has also created an ever-growing number of LLM chatbots and customized platforms far outpacing the ability of manual blocklists to stay current.
This dynamic landscape underlines a need for an adaptive solution capable of dynamic discovery and temporarily blocking of LLM access during high-stakes assessments without disrupting legitimate educational use. While DNS-based blocking has been effective for static services like ad networks, its rigidity is unsuitable for the fast-evolving AI domain\cite{carr2023monitoring}. To address this gap, we propose AI-Sinkhole, an AI agent-augmented DNS blocking framework capable of discovery, semantic classification, and temporarily blocking of emerging AI chatbot domains. 

Unlike static approaches, AI-Sinkhole framework uses a crawling component to continuously monitor news sites, product showcases, code repositories, and historical query logs to identify potential emerging LLM chatbot domains. It then employs an LLM for semantic classification of the domains, which can be used to update DNS blocklists using systems such as Pi-Hole\cite{pihole2025} accordingly. Focusing on preemptive network-level discovery rather than static rules or packet-level inspection, AI-Sinkhole minimizes false positives and privacy concerns.
The remainder of this paper is organized as follows. Section 2 provides an overview of related work, covering LLMs, their transformative role in education and other domains, challenges and risks, and the inadequacy of traditional detection methods. Section 3 introduces the proposed AI-Sinkhole framework, detailing its architecture and key components. Section 4 presents empirical evaluation results using three state-of-the-art open-weight quantized LLMs and a curated dataset of LLM chatbot domains alongside non-LLM websites. We also present our observations in implementing the system in the hope that it will give insight for related research. Section 5 discusses results and implications. Section 6 concludes the paper reflecting on limitations, and proposing directions for future research.
\section{Related Work}
Various studies have demonstrated how LLMs are revolutionizing information processing with their ability to understand and generate natural language at human-level\cite{ye2025llms4all}. These models have now demonstrated exceptional capability in answering complex questions at expert level. The capability of these models is also continually evolving covering a wide range of domains as evidenced by diverse benchmarks\cite{white2024livebench}. 

A key practical application where LLMs have gained popularity is education, where LLMs are redefining traditional pedagogical methods with personalized learning \cite{wen2024ai}. Their ability to synthesize complex information and adapt to diverse learning styles has positioned them as indispensable educational tools that serve as dynamic textbooks\cite{sosnovsky2025intelligent} or research assistants\cite{schmidgall2025agent}, enabling students to explore subjects through interactive and context-aware content. 

Focusing on the student perception and effect of LLMs in essay writing Bernabei et al \cite{bernabei2023students} performed a case study on engineering education. They found that LLMs improved student essay writing. They also found that students' understanding of the topic during oral presentations was improved, showing benefits of LLM assistance. However, they report that many student submissions did not sufficiently rework nor discuss the text produced by ChatGPT, sometimes resulting in contradictory paragraphs. Furthermore, their evaluation of 13 AI detection tools resulted in negative outcome which showed their unreliability in identifying AI generated text, this is in line with previous findings\cite{weber2023testing}.

In a mixed study on the integration LLMs into higher education, Zhang et al. found that LLMs are beneficial for personalized learning and enhancing interaction, thus improving classroom engagement\cite{zhang2024reflections}. Their result also found that professors are concerned with data privacy issues, while students expressed worries about over-reliance on AI would affect their independent learning. The authors concluded that the integration of LLMs into higher education must be managed carefully. To address privacy challenges and the need for LLMs with pedagogically valid responses, Fawzi et al \cite{fawzi2025scribe} proposed SCRIBE, a framework to develop smaller distilled LLMs that can be used as privacy-preserving educational assistants. Even though their experiment demonstrate superior results with tool calling capability, their methodology suggests each specific domain might require its own LLM.

In a tangentially related literature, researchers have studied the varying degree of unsuitability of different LLMs in tasks  such as in tutoring, Macina et al.\cite{macina2025mathtutorbench}. In such scenarios, AI-Sinkhole can be used as an enforcement mechanism to block access to erroneous model providers.

Though adoption of LLMs in educational settings shows great promise, researchers have raised concerns about eminent over-reliance on these tools\cite{gerlich2025ai}. After finding a negative correlation between AI cognitive offloading and critical thinking, Gerlich \cite{gerlich2025ai} highlighted the potential cognitive costs of AI tool reliance, calling for critical AI engagement strategies in education. Multiple researchers have also studied potential challenges and countermeasures of LLM misuse in education, \cite{cotton2024chatting,perkins2023game,borges2024could}. Solutions were proposed to detect anomalous behavior of examinees  such as based on examinee head pose estimation and eye-gaze estimate\cite{singh2023eagle,potluri2022comprehensive}, we consider such methods uncomfortable and privacy invasive.

The objective of out study is how we can temporarily block dynamically emerging LLM chatbot services during proctored academic assessment. In this regard we found network level blocking of LLM serving websites most suitable because of their low overhead and ease of integration to existing network-wide systems such as pi-hole\cite{uramova2024contribution}. However, the dynamic nature of the LLM and generative AI ecosystem limits their effectiveness\cite{carr2023monitoring}.
Our proposed framework addresses this gap leveraging LLM augmented semantic URL classification using multiple pieces of a website. The detailed methodology is described in the following section.
\begin{figure}[ht!]
  \centering
  \includegraphics[width=\linewidth, trim=0 4 4 4,clip]{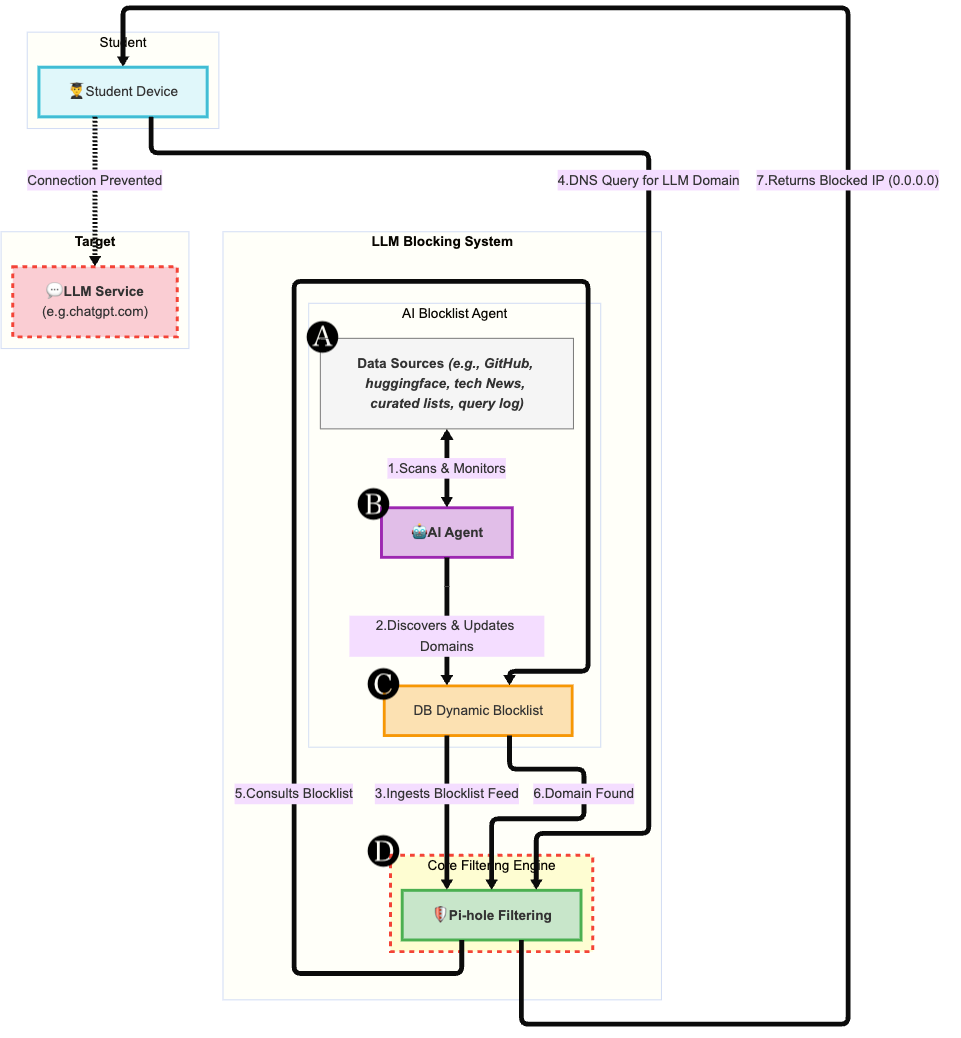}
  \caption{AI-Sinkhole Framework}
  \label{archful}
\end{figure}
\section{Architecture and workflow of AI-Sinkhole Framework} 
\subsection{Architecture of AI-Sinkhole Framework} 
In this section we introduce the AI-Sinkhole framework which is a multi-layered system composed of four main components (see Figure~\ref{archful}): Website discovery and crawling component (box A in the diagram), LLM-based website classification component (box B), dynamic blocklist component(box C), and filtering engine component (box D). Each component is described as follows:

\begin{enumerate}[label=(\Alph*)]
    \item \textbf{Website Discovery and Crawling Component:} \\ 
    This component handles the intelligence-gathering part whose main goal is to continuously scan and monitor a set of assigned data sources and identify a potential list of website URLs. The potential data sources can be news cites (e.g. Hacker News, or TechCrunch), product show cases (e.g. product hunt, Indie hackers) or AI model and code repositories (e.g. Github or Hugging Face). Another very important data source we recommend is recent query logs of user accesses (e.g. Pi-hole DNS query logs) to get an anonymized list of potential URLs that are already being frequented by users. The rationale behind this is that users tend to use the services and UIs they are already accustomed, as suggested by Jakob's law\cite{yablonski2024laws}, with the credentials and account balances they already have.
    
    After a list of potential website URLs is identified this component crawls each website and prepares a JSON file per website that will be the input for module B. The file consists of three pieces of information. The first piece is the URL itself, during our experiments we have observed that LLMs can infer a website's purpose from its URL, for example the following is part of the LLM output from one of our experiments \textit{``The domain name 'trychatgp.com' suggests it it is intended to provide a chat interface for a generative AI service ..."}. The second piece is a dynamically generated website metadata inclduing its title, description, and keywords usually found in the $<head>$ part of a webpage. The next and most important part is the webpage content. Given the dynamic nature of today's websites built with serveral types of frameworks and libraries, it is difficult to create crawlers for each individual website that collect meaningful content from each website. Thankfully, crawler projects like Crawl4AI\cite{crawl4ai2024} can now extract a markdown summary document from a website that can be a direct input to an LLM. We recommend Crawl4AI for this component of our  proposed framework.
    
    \item \textbf{LLM-based Website Classification Component:}\\
    Once the website information is generated, this component handles the classification task using AI agent. The task involves leveraging the website information as input and deciding wether it should be added to the blocklist. Given the versatile nature of the ecosystem including emergence of different terminologies, availability in several languages and geographic regions, and existence of niche and less popular websites, we aim to leverage an AI agent using state of the art LLMs instead of developing a classifier in-house. Note that LLMs have been effective in similar tasks\cite{bozzolan2025llm}. We created a prompt that specifically instructs the AI agent to generate a JSON response of a format \textit{``\{verdict":"YES"/"NO", ``"reason":``$<$explanation for verdict"$>$\}} based on the supplied website information. Adding \textit{reason} together with response is to aid in explaining why the LLM made that decision, this helps a human inspector evaluate decisions\cite{singh2025model}.
    We performed our experiment using Ollama\cite{Ollama}, a local LLM platform, testing three quantized state-of-the-art LLMs, obtaining promising results as discussed in Evaluation and Results section.
    
    \item \textbf{Dynamic blocklist Component:}\\
    This component of our proposed framework handles processing of discovered AI chatbot domains from component B (i.e. those results with $'YES'$ Verdict), and registering them in the block list database of the DNS server with a comment tag $'AI-sinkhole'$. This component also handles whitelisting of the marked domain list (i.e. those having $'AI-sinkhole'$ tag) after a time period (hence allowing temporary blocking). 
    
    For a scenario of multiple networks, this component is designed to be implemented as a centralized list management service, conceptually similar to an adblock subscription model. As new domains are identified, they will be periodically pushed to this central service. Pi-hole installations across various networks can then periodically poll this endpoint to automatically retrieve and apply the latest AI-Sinkhole domain list.
   
    \item \textbf{Filtering Engine Component:}\\
    This component hosts the main functionality that handles the actual blocking of user queries using Pi-Hole\cite{pihole2025}. Pi-hole is DNS server with integrated blocking logic. It is widely used by the privacy community as a network-wide DNS sinkhole for ads and trackers without the need for client-side software. DNS sinkholing is one of cybersecurity measures practiced to mitigate cyberattacks, for example by blocking communication to a botnet command and control center \cite{jung2017efficient}.

    The basic operation of Pi-hole is as follows: When a user sends a DNS request (e.g. chatgpt.com's IP address) Pi-hole checks its blocklist, if the domain is in the list Pi-hole "sinks" the request responding with a non-routable address (e.g. 0.0.0.0), otherwise it serves as normal DNS server. It should be noted that for this system to work Pi-hole should be set as a network-wide DNS server (i.e. DNS server of each device under administration should be configured to Pi-hole's address).
    For the purpose of AI-sinkhole, the default blocklist of thousands of URLs that comes with Pi-hole is not required (it can be removed via the web interface, steps available on AI-sinkhole's Github page) as that list is not the goal of this project.
\end{enumerate}
\vspace{-1cm}
\subsection{Workflow of AI-Sinkhole Framework}
The complete lifecycle of identifying an AI chatbot and blocking it can be explained following the diagram on Figure~\ref{archful}. Initially, the system continuously monitors set of data sources (step 1 in the diagram), if new domains are discovered they will be added to the dynamic blockist (step 2), the core filtering engine will then ingest these newly discovered domains (step 3). The filtering engine (pi-hole) then uses the updated list of blocklists while preserving access to legitimate educational tools. When a user's browser tries to connect to an LLM service, a DNS query is sent to the core filtering engine (step 4), the engine consults its cache and its blocklist (step 5), if the domain matches a known chatbot service (step 6), the engine redirects the request to a sinkhole IP (e.g., 0.0.0.0), effectively preventing the connection establishment (Step 7).

\section{Evaluation and Results}
\subsection{Experiment Setup}
  To evaluate the efficacy of the proposed framework we performed an experiment targeting the two core components of the framework: (1) the AI agent’s classification performance for identifying LLM chatbot services, and (2) the framework’s ability to implement temporary, dynamic DNS-based blocking for emerging LLM domains.
\begin{figure}[t!]
  \centering
  \includegraphics[width=0.75\linewidth]{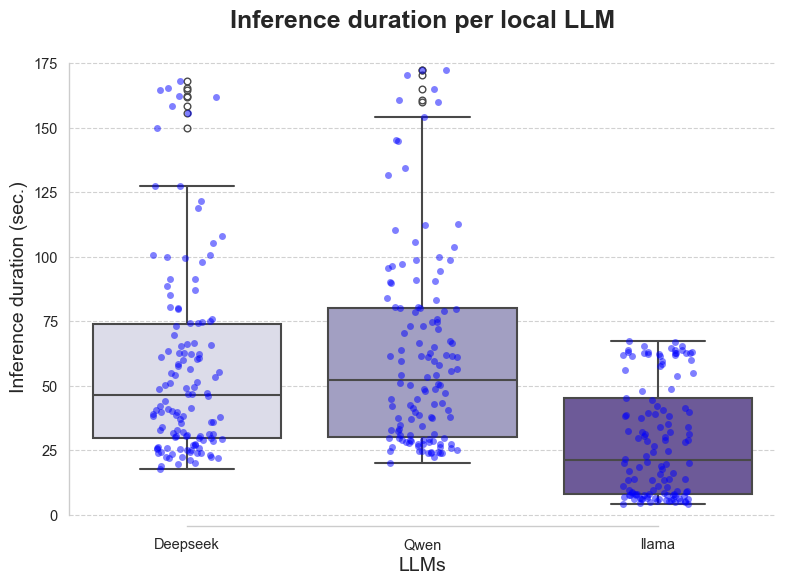}
  \caption{Inference latency of the three quantized LLMs}
  \label{fig:model-latency}
\end{figure}
  \subsubsection{Dataset}
 We curated a dataset of known website URLs split into positive (LLM chatbot services) and negative (non-LLM websites) classes.
The positive class included domains hosting LLM chatbots (e.g., chatgpt.com, meta.ai, etc.) or aggregator services (e.g. poe.com). We curated a known list of 63 such websites collected from news sites, blogs, and GitHub lists (details available on GitHub).

To balance this list with a negative class we curated a list of 63 known websites that do not serve LLM. To diversify this negative class we composed this list from websites of education institutions, online tutorial websites, software documentation pages, and government and NGO websites. We also tested the effectiveness of detecting false positives by including webpages that discuss LLMs and chatbots. We also evaluated model capabilities to classify websites in different languages, we included Spanish, German, Mandarin, Arabic, French, Portuguese, and English. A total of 126 websites were used to test crawling capability and classification performance on these websites.

  \subsubsection{LLM models and implementation details}
The experiment was conducted using:
\begin{itemize}

\item Three quantized LLMs (LLama 3, deepseek-r1, and qwen 3) via Ollama (Version 0.12) on a laptop running Apple M2 processor with GPU acceleration for classification tasks, Table~\ref{table:model-comparison} summarizes parameters and quantization of the models.
\item Crawl4AI (v0.7) library for metadata and web content summarization.
\item Pi-Hole as the DNS sinkhole mechanism deployed on docker on a  windwos a machine, with $'AI-sinkhole'$  blocklist enabled-disabled-enabled for 12 hours to simulate temporary blocking ability.
\item Webbrowser python library for simulating client web-browsing running on a linux VM.
\end{itemize}
A prompt was crafted after multiple iterations of interactions with the LLMs by examining both the results we receive and the explanations behind them. The prompt template used for this experiment along with a sample response is shown on Figure~\ref{fig:prompt}.
To validate the framework’s dynamic blocking functionality, we added the positive classes  to a blocklist with $``AI-Sinkhole"$ comment tag and simulated iterative browsing of both positive and negative classes at three time intervals: before enabling $``AI-Sinkhole"$, while it is enabled, and after disabling it to confirm temporary blocking.
\begin{figure}[t!]
    \centering
    \begin{tcolorbox}[
    colback=purple!5,colframe=blue!40!purple,title=Prompt template, halign title=center, fonttitle=\bfseries\Large, box align=top]
    \textbf{\colorbox{cyan!40!white}{User prompt:} }
    I need to classify a website based on its primary function. The following is partial information extracted from the website. \{weblog\_data\}
    My Question: Does this website (i.e. the domain name given in above URL) provide a general-purpose, generative AI chat service for answering a wide variety of questions? Please check if it meets these criteria: Functions like ChatGPT, Grok, meta.ai, or Gemini: It can understand and answer a broad range of general knowledge questions (e.g., "Explain photosynthesis", "Help me solve this algebra problem", "Write an essay outline"), or it serves as frontend chat interface for accessing such LLMs ... Please answer with ``Yes" or ``No" and provide a brief explanation of what the website is, based on the provided info. the format of your answer should only contain a JSON formatted response as \{``verdict": answer, ``reason": reason\}\\
    \textbf{\colorbox{cyan!40!white}{Sample LLM response:} }
                \{"verdict": "No","reason": "The website appears to be an online education platform offering courses and programs in various fields, including artificial intelligence. While it mentions LLMs (Large Language Models) and AI agents, its primary function is not a general-purpose chat service for answering a wide variety of questions."\}

    \end{tcolorbox}
    \caption{Prompt template used in the experiment with a sample response}
    \label{fig:prompt}
\end{figure}
\vspace{-1cm}
\subsection{Results}
\begin{figure}[t]
     \centering
       \begin{tabular}{ll}
           \includegraphics[width=0.48\linewidth]{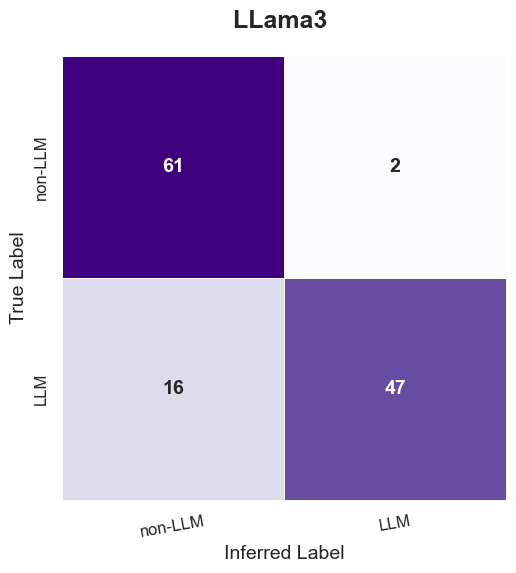} &
           \includegraphics[width=0.48\linewidth]{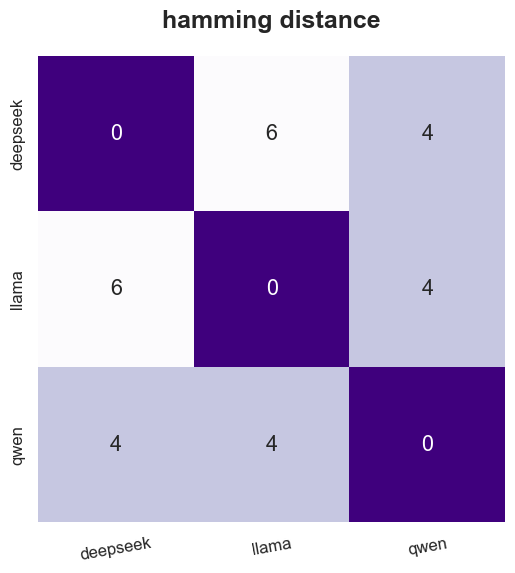}
   \end{tabular}
   \caption{LLM comparison. Left: Confusion matrices comparing LLM decisions (LLama 3) with the ground truth. Right: hamming distance showing similarity of model decisions.}
   \label{fig:cofusing-hamming}
     \vspace*{-0.18in}
   \end{figure}

The proposed AI-Sinkhole framework demonstrated robust performance in both classifying LLM chatbot services and implementing dynamic DNS-based blocking. A total of 126 websites (63 LLM chatbot services and 63 non-LLM services) were evaluated for classification accuracy, F1 score and Matthews Correlation Coefficient (MCC as shown on eq~\ref{eq:mcc_formula}) with metrics summarized in Table~\ref{table:model-comparison}. The three quantized LLMs: LLama 3, DeepSeek-R1, and Qwen 3 showed a very good performance and similar results (as indicated by the hamming distance on Figure~\ref{fig:cofusing-hamming}, with LLama 3 achieving marginally higher F1-score (0.839) while Qwen 3 and DeepSeek-R1 showing a slighly higher score on MCC (MCC = 0.738)). In comparison with precision, the classifiers' recall was less strong across all models, which showed models' tendency to give higher False negatives. Analysis of explanations gave insights to some of the decisions including: websites detecting our automated crawler, LLMs mis-classifying some aggregator websites, LLMs correctly classifying companies that have recently changed their business (e.g. you.com, which preciously was an AI search engine now focusing on enterprise solutions).

To assess the framework’s temporal blocking capability, we simulated user browsing of LLM and non-LLM domains for 12 hours at three equal intervals: pre-blocklist (baseline), during AI-Sinkhole activation, and post-deactivation. During activation, the DNS sinkhole successfully blocked $\textbf{100}$\% of LLM domains ($\textbf{63}$/63) while maintaining access to all non-LLM services, The blocking duration was sustained for 4 hours, and after deactivation, the system reverted to unblocked access, confirming the framework’s temporary and dynamic capability. Cross-lingual evaluation also validated the framework’s adaptability. When classifiers were tested on websites in the above multiple languages they showed language agnostic ability, correctly understanding the prompt in English to classify websites in other languages. The results demonstrate the viability of the proposed framework for managing access to LLM chatbot services during academic evaluations, balancing performance and low operational overhead.

    \begin{equation}
        \text{MCC} = \frac{\text{TP} \times \text{TN} - \text{FP} \times \text{FN}}{\sqrt{(\text{TP} + \text{FP})(\text{TP} + \text{FN})(\text{TN} + \text{FP})(\text{TN} + \text{FN})}}
        \label{eq:mcc_formula}
    \end{equation}
    
\subsection{Observations}
 The findings from this study highlight a few critical insights into the emerging practical application of LLMs as classifiers which can be extended to other domains such as cybersecurity (e.g., for dynamic threat detection). First, \textbf{cross-language versatility emerged as a key strength of LLMs as classifiers}: the AI-Sinkhole framework demonstrated versatility in classification performance across Spanish, German, Mandarin, Arabic, French, Portuguese, and English with an average F1-score above 0.83. This adaptability demonstrates the potential of such LLMs to address the globalized nature of the Web avoiding the need for fine-tuning.\\
Second, the \textbf{importance of incorporating explanations in classification responses} was evident in the framework’s ability to distinguish between the two classes. By explicitly incorporating explanations into the classification result, we were able to get the explanations behind each classification which helped us improve our prompt  particularly in cases where domain descriptions or metadata lacked clear indicators of LLM service hosting. This highlights the necessity of integrating explainability mechanisms into AI systems to ensure transparency and trustworthiness in high-stakes applications \cite{singh2025model}.\\
Third, the \textbf{LLM’s capability to identify niche or low-profile chatbot services} demonstrated the framework’s adaptability to emerging threats. Unlike traditional blocklists reliant on static domain lists, the AI agent successfully detected recently launched or obscure LLM services, even when their URLs lacked explicit keywords. This ability to recognize subtle behavioral or semantic cues rather than solely surface-level indicators positions such LLM based frameworks as superior tools for dynamic classification tasks.\\
Finally, \textbf{a negligible trade-off between performance and inference latency was observed}: as shown on Figure~\ref{fig:model-latency} ``Thinking" models ( DeepSeek- R1, Qwen 3) exhibited approximately 2X inference latency compared to a lighter, non-reasoning LLama 3 model. 
Using ``Thinking" models to this classification task offered minimal performance gains while significantly increasing inference latency.
High inference latency limits real-time scalability for high-traffic networks suggesting the need for  multi-dimensional evaluation of models as discussed on \cite{farjad2025secure}. Such evaluation can be leveraged to build hybrid models that balance fast inference for routine tasks with reasoning for critical decisions.

\section{Discussion}
The results of this study validate the feasibility of leveraging large language models (LLMs) as dynamic classifiers for detecting and blocking emergent LLM chatbot services during academic evaluations, addressing a critical gap in the domain. The cross-lingual performance of the proposed AI-Sinkhole framework highlights the potential of quantized local LLMs  in classification tasks extending to other domains such as globalized cybersecurity threats without extensive language-specific fine-tuning. This adaptability is particularly relevant in educational ecosystems where multilingual content and international student populations are common, as it reduces the operational burden of maintaining separate classifiers for each language.
A key distinction of the AI-Sinkhole framework lies in our integration of explanation with classification, which significantly improves the accuracy of distinguishing between LLM chatbots and domain content discussing AI tools.
This capability to extract explanatory justifications for classification decisions enhances model transparency and provides actionable insights for refining prompts and improving system explainability which is a critical need in high-stakes applications\cite{singh2025model,ramsey2023toward}. We also showed a trade-off between model ``thinking" capability and inference latency which presents a practical challenge for real-time deployment necessitating multi-dimensional model evaluation\cite{farjad2025secure}.
Finally, the limitations of current AI detection systems to adapt to rapidly evolving LLM platforms underscores the urgency of innovative frameworks like AI-Sinkhole as a technical solution and also the need for revisioning of academic evaluation methodologies. While proposing alternative pedagogical evaluations (e.g. Oral exams in Italy \cite{melcarne2025oral}) is outside of the scope of this paper \textbf{we emphasize that it is essential to balance embracing the benefits of AI-powered learning with the need to maintain goals of academic assessment}.

\begin{table}[t!]
\centering
\caption{LLM model specifications and prediction results}
\label{table:model-comparison}
\begin{tabular}{ccccccc}
\toprule
 & \multicolumn{2}{c}{\textbf{LLM model}} & \multicolumn{3}{c}{\textbf{Prediction}} \\ 
\cmidrule(lr){2-3} \cmidrule(lr){4-6}
 & \textbf{params} & \textbf{quantization} &\textbf{accuracy} & \textbf{F1} &MCC\\ 
\midrule
\textbf{LLAMA}    & 8.0B   & Q4\_0      & 0.857     & 0.839& 0.732 \\ 
\textbf{Deepseek} & 8.2B    &  Q4\_K\_M & 0.857    &  0.836&  0.738 \\ 
\textbf{Qwen}     &  8.2B & Q4\_K\_M    &  0.857   & 0.836& 0.738 \\ 
\bottomrule
\end{tabular}
\end{table}
\begin{figure*}[h!]
  \centering
  \includegraphics[width=0.75\linewidth]{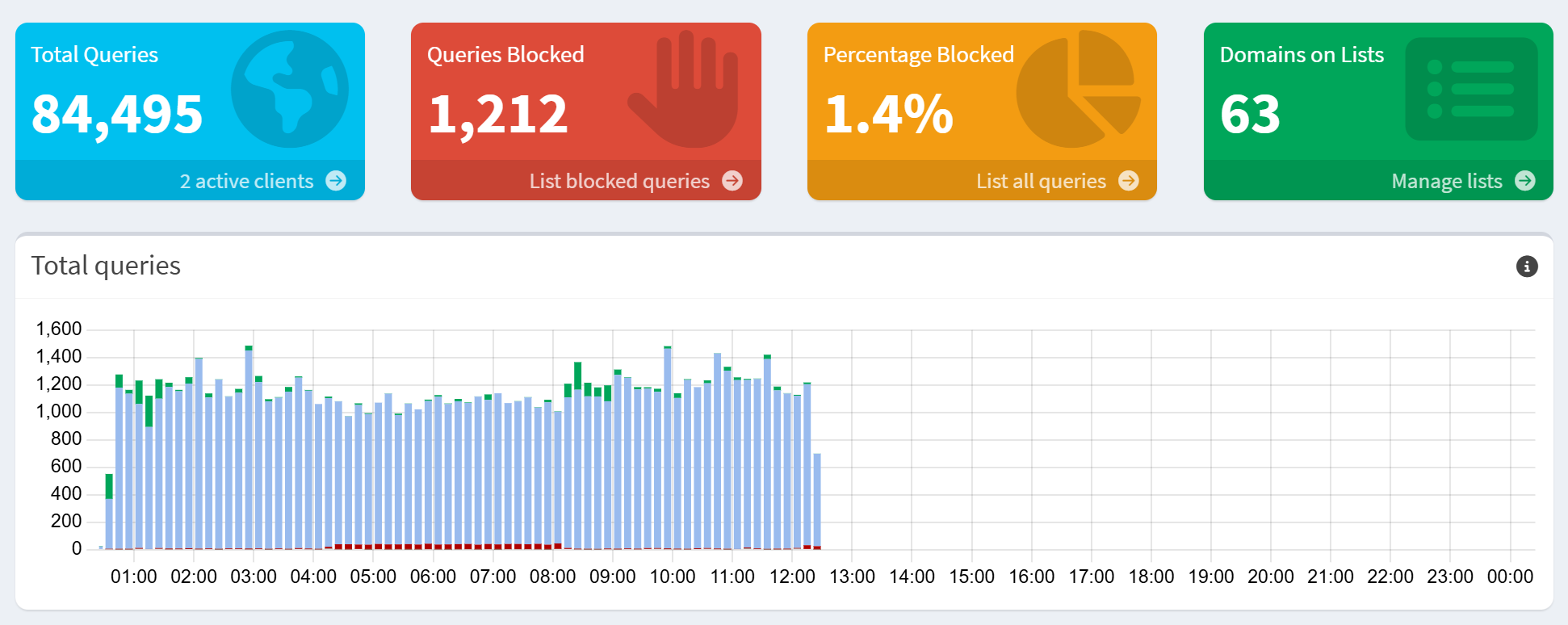}
  \caption{Pi-hole dashboard showing AI-Sinkhole deployment showing 63 AI domains on the blocklist.}
  \label{fig:pihole-demo}
\end{figure*}
\vspace{-0.5cm}
\section{Conclusion and Future Work}
This study introduces the AI-Sinkhole framework integrating quantized LLMs with crawling, explainable classification, and dynamic DNS blocking. The framework addresses the limitations of existing systems, offering a scalable and adaptable solution to the growing challenge of AI misuse in education.
The experimental results demonstrate encouraging contribution of the framework toward balancing academic integrity with practical AI integration. Future work should focus on expanding the framework’s deployment beyond DNS, optimizing inference efficiency for large-scale deployments, and exploring its applicability in broader cybersecurity contexts. By addressing the limitations of current AI detection systems, this framework offers a promising pathway to educational environments that integrate AI technologies.
\begin{acknowledgements}
This work is partially supported by a grant award from the United
States National Science Foundation (NSF) – NSF Grant ID
2234554.
\end{acknowledgements}
\bibliographystyle{spphys}  
\bibliography{refs}

@String{Computing = "Computing" }

@String{Academic = "Academic Press" }

@String{Springer = "Springer-Verlag" }

@article{yang2025agentic,
  title={Agentic web: Weaving the next web with ai agents},
  author={Yang, Yingxuan and Ma, Mulei and Huang, Yuxuan and Chai, Huacan and Gong, Chenyu and Geng, Haoran and Zhou, Yuanjian and Wen, Ying and Fang, Meng and Chen, Muhao and others},
  journal={arXiv preprint arXiv:2507.21206},
  year={2025}
}

@article{schmidgall2025agent,
  title={Agent laboratory: Using llm agents as research assistants},
  author={Schmidgall, Samuel and Su, Yusheng and Wang, Ze and Sun, Ximeng and Wu, Jialian and Yu, Xiaodong and Liu, Jiang and Moor, Michael and Liu, Zicheng and Barsoum, Emad},
  journal={arXiv preprint arXiv:2501.04227},
  year={2025}
}

@article{sosnovsky2025intelligent,
  title={Intelligent textbooks},
  author={Sosnovsky, Sergey and Brusilovsky, Peter and Lan, Andrew},
  journal={International Journal of Artificial Intelligence in Education},
  pages={1--20},
  year={2025},
  publisher={Springer}
}

@inproceedings{wen2024ai,
  title={AI for education (AI4EDU): Advancing personalized education with LLM and adaptive learning},
  author={Wen, Qingsong and Liang, Jing and Sierra, Carles and Luckin, Rose and Tong, Richard and Liu, Zitao and Cui, Peng and Tang, Jiliang},
  booktitle={Proceedings of the 30th ACM SIGKDD Conference on Knowledge Discovery and Data Mining},
  pages={6743--6744},
  year={2024}
}

@article{white2024livebench,
  title={Livebench: A challenging, contamination-free llm benchmark},
  author={White, Colin and Dooley, Samuel and Roberts, Manley and Pal, Arka and Feuer, Ben and Jain, Siddhartha and Shwartz-Ziv, Ravid and Jain, Neel and Saifullah, Khalid and Naidu, Siddartha and others},
  journal={arXiv preprint arXiv:2406.19314},
  volume={4},
  year={2024}
}

@article{macina2025mathtutorbench,
  title={Mathtutorbench: A benchmark for measuring open-ended pedagogical capabilities of llm tutors},
  author={Macina, Jakub and Daheim, Nico and Hakimi, Ido and Kapur, Manu and Gurevych, Iryna and Sachan, Mrinmaya},
  journal={arXiv preprint arXiv:2502.18940},
  year={2025}
}

@article{zhang2024reflections,
  title={Reflections on Enhancing Higher Education Classroom Effectiveness Through the Introduction of Large Language Models},
  author={Zhang, Xiaoming and Zhang, Xiaoli and Liu, Heping},
  journal={J Mod Educ Res},
  year={2024},
  publisher={Innovation Forever Publishing Group Limited}
}

@inproceedings{fawzi2025scribe,
  title={SCRIBE: Structured Chain Reasoning for Interactive Behaviour Explanations using Tool Calling},
  author={Fawzi, Fares and Swamy, Vinitra and Glandorf, Dominik and Nazaretsky, Tanya and K{\"a}ser, Tanja},
  booktitle={Proceedings of the 2025 Conference on Empirical Methods in Natural Language Processing},
  pages={29273--29298},
  year={2025}
}

@article{ye2025llms4all,
  title={LLMs4All: A Systematic Review of Large Language Models Across Academic Disciplines},
  author={Ye, Yanfang and Zhang, Zheyuan and Ma, Tianyi and Wang, Zehong and Li, Yiyang and Hou, Shifu and Sun, Weixiang and Shi, Kaiwen and Ma, Yijun and Song, Wei and others},
  journal={arXiv e-prints},
  pages={arXiv--2509},
  year={2025}
}

@article{bernabei2023students,
  title={Students’ use of large language models in engineering education: A case study on technology acceptance, perceptions, efficacy, and detection chances},
  author={Bernabei, Margherita and Colabianchi, Silvia and Falegnami, Andrea and Costantino, Francesco},
  journal={Computers and Education: Artificial Intelligence},
  volume={5},
  pages={100172},
  year={2023},
  publisher={Elsevier}
}

@inproceedings{singh2023eagle,
  title={Eagle Eye: Enhancing Online Exam Proctoring Through AI-powEred Eye Gaze Detection},
  author={Singh, Jagendra and Mishra, Amit Kumar and Chopra, Leena and Agarwal, Gunjan and Diwakar, Manoj and Singh, Prabhishek},
  booktitle={International Conference on Electrical and Electronics Engineering},
  pages={173--185},
  year={2023},
  organization={Springer}
}

@inproceedings{potluri2022comprehensive,
  title={A comprehensive survey on the AI based fully automated online proctoring systems to detect anomalous behavior of the examinee},
  author={Potluri, Tejaswi and Sistla, Venkatarama Phani Kumar},
  booktitle={2022 International Conference on Recent Trends in Microelectronics, Automation, Computing and Communications Systems (ICMACC)},
  pages={407--411},
  year={2022},
  organization={IEEE}
}

@article{gerlich2025ai,
  title={AI tools in society: Impacts on cognitive offloading and the future of critical thinking},
  author={Gerlich, Michael},
  journal={Societies},
  volume={15},
  number={1},
  pages={6},
  year={2025},
  publisher={Multidisciplinary Digital Publishing Institute}
}

@article{weber2023testing,
  title={Testing of detection tools for AI-generated text},
  author={Weber-Wulff, D and et al.},
  journal={International Journal for Educational Integrity},
  volume={19},
  number={1},
  pages={1--39},
  year={2023},
  publisher={Springer}
}

@article{cotton2024chatting,
  title={Chatting and cheating: Ensuring academic integrity in the era of ChatGPT},
  author={Cotton, Debby RE and Cotton, Peter A and Shipway, J Reuben},
  journal={Innovations in education and teaching international},
  volume={61},
  number={2},
  pages={228--239},
  year={2024},
  publisher={Taylor \& Francis}
}

@article{perkins2023game,
  title={Game of Tones: Faculty detection of GPT-4 generated content in university assessments},
  author={Perkins, Mike and Roe, Jasper and Postma, Darius and McGaughran, James and Hickerson, Don},
  journal={arXiv preprint arXiv:2305.18081},
  year={2023}
}

@article{borges2024could,
  title={Could ChatGPT get an engineering degree? Evaluating higher education vulnerability to AI assistants},
  author={Borges, Beatriz  and others},
  journal={Proceedings of the National Academy of Sciences},
  volume={121},
  number={49},
  pages={e2414955121},
  year={2024},
  publisher={National Academy of Sciences}
}

@misc{carr2023monitoring,
  title={Monitoring Malicious DNS Queries: An Experimental Case Study of Utilising the National Cyber Security Centre’s Protective DNS within a UK Public Sector Organisation},
  author={Carr, Andrew and Alam, Abu and Allison, Jordan},
  year={2023}
}

@inproceedings{uramova2024contribution,
  title={Contribution to Safer Internet for Children at School},
  author={Uramova, Jana and Segec, Pavel and Moravc{\'\i}k, Marek},
  booktitle={2024 International Conference on Emerging eLearning Technologies and Applications (ICETA)},
  pages={1--7},
  year={2024},
  organization={IEEE}
}

@incollection{melcarne2025oral,
  title={Oral Exams: Balancing Assessment Integrity with Fairness--Lessons from Italian Medical Education},
  author={Melcarne, R and Yucel, S and Vitale, S and De Benedictis, D and Leopardi, E and Violani, C and Familiari, G and Maranghi, M and D'Amati, Giulia and others},
  booktitle={AMEE 2025},
  year={2025},
  publisher={AMEE}
}

@inbook{singh2025model,
author = {Singh, Vidit and Kassa, Yonas and Kale, Akshay and Ricks, Brian and Gandhi, Robin},
year = {2025},
month = {05},
pages = {316-326},
title = {Model-Cart: A Machine Learning Meta-Framework with Explainability and Human-in-the-Loop},
isbn = {978-3-031-89062-8},
publisher={Springer},
doi = {10.1007/978-3-031-89063-5_27}
}

@book{yablonski2024laws,
  title={Laws of UX},
  author={Yablonski, Jon},
  year={2024},
  publisher={" O'Reilly Media, Inc."}
}

@inbook{kassa2025online,
author="Kassa, Yonas",
title="Online Advertising is a Regrettable Necessity: On the Dangers of Pay-Walling the Web",
booktitle="The 22nd International Conference on Information Technology-New Generations (ITNG 2025)",
year="2025",
publisher="Springer Nature Switzerland",
address="Cham",
pages="615--624",
isbn="978-3-031-89063-5",
doi = {10.1007/978-3-031-89063-5_53}

}

@misc{crawl4ai2024,
  author = {UncleCode},
  title = {Crawl4AI: Open-source LLM Friendly Web Crawler \& Scraper},
  year = {2024},
  publisher = {GitHub},
  journal = {GitHub Repository},
  howpublished = {\url{https://github.com/unclecode/crawl4ai}},
}

@article{bozzolan2025llm,
  title={LLM-Assisted Web Measurements},
  author={Bozzolan, Simone and Calzavara, Stefano and Cazzaro, Lorenzo},
  journal={arXiv preprint arXiv:2510.08101},
  year={2025}
}

@misc{Ollama,
  title = {{Ollama}: Large Language Models Locally},
  author = {{Ollama team}},
  howpublished = {\url{https://ollama.com/}},
  year = {2025} 
}

@misc{pihole2025,
  title = {{Pi-hole}: Network-wide ad blocking},
  author = {The Pi-hole Project},
  howpublished = {\url{https://pi-hole.net}},
  note = {Accessed: 2025-11-13},
  year = {2025}
}

@article{jung2017efficient,
  title={Efficient malicious packet capture through advanced DNS sinkhole},
  author={Jung, Hyun Mi and Lee, Haeng Gon and Choi, Jang Won},
  journal={Wireless Personal Communications},
  volume={93},
  number={1},
  pages={21--34},
  year={2017},
  publisher={Springer}
}

@inproceedings{farjad2025secure,
author = {Farjad, Sheikh Muhammad and Patllola, Sandeep Reddy and Kassa, Yonas and Grispos, George and Gandhi, Robin},
title = {Secure Edge Computing Reference Architecture for Data-driven Structural Health Monitoring: Lessons Learned from Implementation and Benchmarking},
year = {2025},
isbn = {9798400712777},
publisher = {Association for Computing Machinery},
address = {New York, NY, USA},
url = {https://doi.org/10.1145/3696673.3723074},
doi = {10.1145/3696673.3723074},
booktitle = {Proceedings of the 2025 ACM Southeast Conference},
pages = {145–154},
numpages = {10},
location = {Southeast Missouri State University, Cape Girardeau, MO, USA},
series = {ACMSE 2025}
}

@inproceedings{ramsey2023toward,
  title={Toward interactive visualizations for explaining machine learning models},
  author={Ramsey, Ashley and Kale, A and Kassa, Y and Gandhi, Robin and Ricks, Brian},
  booktitle={Proceedings of the Information Systems for Crisis Response and Management Conference, Omaha, NE, USA},
  pages={28--31},
  year={2023}
}
\end{document}